# Discrete Teissier distribution: properties, estimation and application

[1]Bhupendra Singh, [2]Varun Agiwal, [3]Ravindra Pratap Singh, [1]Abhishek Tyagi[*]

[1]Department of Statistics, Chaudhary Charan Singh University, Meerut-250004, India

[2]Indian Institute of Public Health, Hyderabad, Telangana, India.

[3]Department of Statistics, Central University of Rajasthan, Rajasthan-305817, India.

Corresponding author E-mail: abhishektyagi033@gmail.com

## Abstract

In this article, a discrete analogue of continuous Teissier distribution is presented. Its several important distributional characteristics have been derived. The estimation of the unknown parameter has been done using the method of maximum likelihood and the method of moment. Two real data applications have been presented to show the applicability of the proposed model.

**Keywords:** COVID-19, Classical estimation, Discrete Teissier distribution, method of maximum likelihood, moment estimation.

**MSC:** 60E05; 62F10.

## 1. Introduction

The concept of discretization generally arises from the field of survival/reliability analysis when it becomes impossible or inconvenient to measure the life length of a product/unit on a continuous scale. Few real examples where lifetime is to be recorded on a discrete scale rather than on continuous are the survival times for those suffering from the diseases like lung cancer or period from remission to relapse may be recorded as the number of days, the number of the cycle before the first failure when device work in cycle, the number of to and fro motions of a pendulum or spring device before resting, the number of times the device is switched on/off, modelling probability distribution of count data etc. These are some practical situations that have to catch the eyes of many researchers and hence motivate them to find more plausible discrete distributions to model discrete data arising from various real-life situations.

Discretization of the existing continuous distribution can be done using different methodologies [see Chakraborty, 2015]. Out of these, one widely used methodology is described as follows:

If the underlying random variable $Y$ has the survival function (SF) $S_Y(y) = P(Y \geq y)$ then the random variable $X = [Y]$, largest integer less than or equal to $Y$ will have the probability mass function (PMF)

$$\begin{aligned} P(Y = y) &= P(y \leq X \leq y+1) \\ &= S_X(y) - S_X(y+1); \ y = 0, 1, 2, 3, \ldots \end{aligned} \quad (1)$$

One of the important virtue of this methodology is that the developed discrete distribution retains the same functional form of the SF as that of its continuous counterpart. Due to this feature, many reliability characteristics of the distribution remain unchanged.

Over the last two decades, this approach has gotten a lot of attention. Using this technique, Roy (2003) gave a discretized version of the normal distribution. Following this, Roy (2004) obtained discrete Rayleigh distribution. A comprehensive analysis of the evolution of the discrete distribution up to 2014 was provided by Chakraborty (2015). Then afterwards, a large number of significant discrete distributions have emerged in the literature. For example, Alamatsaz et al. (2016), Jayakumar and Babu (2018), Tyagi et al. (2019), Tyagi et al. (2020), Eliwa et al. (2020), El-Morshedy et al. (2020), and the references cited therein. Most recently, El-Morshedy et al. (2021a) developed discrete generalized Lindley distribution to model the counts of novel coronavirus cases. El-Morshedy et al. (2021b) gave a discrete analogue of the odd Weibull-G family of distributions. They discussed the classical and Bayesian estimations and showed the applicability of the proposed family to count data sets.

In this paper, we have proposed the discrete analogue of the Teissier model [Teissier, 1934], in the so-called discrete Teissier (DT) distribution using the survival discretization method.

The rest of the article is organized as follows: Section 2 introduces the one-parameter DT distribution. In Section 3 some important distributional and reliability characteristics are studied. In section 4, we estimate the parameter of DT distribution by the method of maximum likelihood and method of the moment. Two real data illustrations are presented in Section 5. Finally, some concluding remarks are given in Section 6.

## 2. Discrete Teissier distribution

If $X$ follows univariate continuous Teissier distribution with parameter $\alpha$ then its probability density function (PDF) and SF can be written as

$$f(x,\alpha) = \alpha(\exp(\alpha x) - 1)\exp(\alpha x - e^{\alpha x} + 1); \alpha > 0, x > 0, \quad (2)$$

$$S(x) = \exp(\alpha x - e^{\alpha x} + 1); \alpha > 0, x > 0. \quad (3)$$

Using the survival discretization approach (1), the discrete Teissier (DT) distribution can be obtained as

$$\begin{aligned} p_y = P[Y = y] &= S_X(y) - S_X(y+1) \\ &= \exp(1)\exp(\alpha y)(\exp(-e^{\alpha y}) - \exp(\alpha - e^{\alpha(y+1)})); y = 0,1,2,\ldots, \alpha > 0. \end{aligned} \quad (4)$$

After re-parametrization $\theta = \exp(\alpha)$, the PMF in (4) can be written as

$$p_y = P[Y = y] = \exp(1)\theta^y(\exp(-\theta^y) - \theta\exp(-\theta^{(y+1)})); y = 0,1,2,\ldots, \theta > 1. \quad (5)$$

The cumulative distribution function (CDF) corresponding to PMF (5) is

$$F(x) = 1 - \theta^{y+1}\exp(1 - \theta^{(y+1)}); y = 0,1,2,\ldots, \theta > 1, \quad (6)$$

## 3. Statistical properties

### 3.1 The Shape of the Probability Mass Function

The PMF plots of the DT distribution for different parametric values are shown in Figure 1.

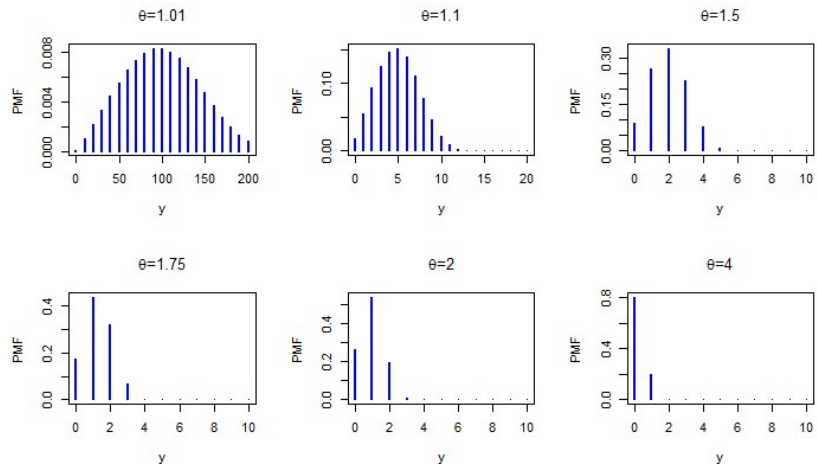

Figure 1. The shapes of PMF of DT distribution for various values of the parameter.

The limiting behavior of DT distribution for various choices of parameters at the boundary points is:

- $\lim_{y \to \infty} p_y = 0$,

- $\lim_{\theta \to 1} p_y = 0$,

- $\lim_{\theta \to \infty} p_y = 1$ for y=0 and $\lim_{\theta \to \infty} p_y = 0$, otherwise.

### 3.2 Moments and related concepts

The $r^{th}$ raw moments say $\varpi'_r$ of the DT distribution can be obtained by using

$$\varpi'_r = E(Y^r) = \sum_{y=0}^{\infty} y^r p_y$$
$$= \exp(1) \sum_{y=1}^{\infty} \sum_{k=0}^{\infty} (-1)^k \frac{(y^r - (y-1)^r)\theta^{(k+1)y}}{\lfloor k}. \tag{7}$$

Using Equation (7), the first four raw moments of the DT distribution are

$$\varpi'_1 = E(Y) = \exp(1) \sum_{y=1}^{\infty} \sum_{k=0}^{\infty} (-1)^k \frac{\theta^{(k+1)y}}{\lfloor k}, \tag{8}$$

$$\varpi'_2 = E(Y^2) = \exp(1) \sum_{y=1}^{\infty} \sum_{k=0}^{\infty} (-1)^k (2y-1) \frac{\theta^{(k+1)y}}{\lfloor k}, \tag{9}$$

$$\varpi'_3 = E(Y^3) = \exp(1) \sum_{y=1}^{\infty} \sum_{k=0}^{\infty} (-1)^k \left(3y^2 - 3y + 1\right) \frac{\theta^{(k+1)y}}{\lfloor k}, \tag{10}$$

$$\varpi'_4 = E(Y^4) = \exp(1) \sum_{y=1}^{\infty} \sum_{k=0}^{\infty} (-1)^k \left(4y^3 - 6y^2 + 4y - 1\right) \frac{\theta^{(k+1)y}}{\lfloor k}. \tag{11}$$

The variance of the DT distribution is

$$V(Y) = \exp(1) \sum_{y=1}^{\infty} \sum_{k=0}^{\infty} (-1)^k (2y-1) \frac{\theta^{(k+1)y}}{\lfloor k} - \left[ \exp(1) \sum_{y=1}^{\infty} \sum_{k=0}^{\infty} (-1)^k \frac{\theta^{(k+1)y}}{\lfloor k} \right]^2.$$

Using the raw moments in (8)-(11), we can easily find the skewness (Sk) and kurtosis (Kur) from the following relations

$$Sk = \frac{\varpi_3' - 3\varpi_2'\varpi_1' + 2(\varpi_1')^3}{(Var(Y))^{3/2}}, \text{ and } Kur = \frac{\varpi_4' - 4\varpi_3'\varpi_1' + 6\varpi_2'(\varpi_1')^2 - 3(\varpi_1')^4}{(Var(Y))^2}, \text{ respectively.}$$

The moment generating function (MGF) for the proposed model is

$$M_Y(t) = E[\exp(ty)] = \sum_{y=0}^{\infty} \exp(ty) p_y$$
$$= 1 + \exp(1)\theta(\exp(t)-1)\sum_{y=1}^{\infty} \exp(-\theta^y).(\theta\exp(t))^{y-1}.$$

The index of dispersion (IOD) in the case of the proposed model is

$$IOD = \frac{Var(Y)}{E(Y)} = \frac{\sum_{y=1}^{\infty}\sum_{k=0}^{\infty}(-1)^k (2y-1)\frac{\theta^{(k+1)y}}{\lfloor k} - \exp(1)\left[\sum_{y=1}^{\infty}\sum_{k=0}^{\infty}(-1)^k \frac{\theta^{(k+1)y}}{\lfloor k}\right]^2}{\sum_{y=1}^{\infty}\sum_{k=0}^{\infty}(-1)^k \frac{\theta^{(k+1)y}}{\lfloor k}}.$$

The coefficient of variation (CV) for DT distribution can be obtained as

$$IOD = \frac{(Var(Y))^{1/2}}{E(Y)} = \frac{\left(\sum_{y=1}^{\infty}\sum_{k=0}^{\infty}(-1)^k (2y-1)\frac{\theta^{(k+1)y}}{\lfloor k} - \exp(1)\left[\sum_{y=1}^{\infty}\sum_{k=0}^{\infty}(-1)^k \frac{\theta^{(k+1)y}}{\lfloor k}\right]^2\right)^{1/2}}{\sum_{y=1}^{\infty}\sum_{k=0}^{\infty}(-1)^k \frac{\theta^{(k+1)y}}{\lfloor k}}.$$

Due to the non-closure form of the above expressions, we use R software to demonstrate these characteristics numerically. Table 1 lists some numerical results of the mean, variance, skewness, kurtosis, IOD, and CV for the DT distribution under different setups of parametric values. From this table, it can be concluded that:

- The mean of the DT distribution decreases when the value of $\theta$ increases.
- From the observed values of skewness, we can conclude that the DT distribution can be used to model positively and negatively skewed data.
- The proposed model is appropriate for modeling leptokurtic and platykurtic data sets.
- The DT distribution can be used to analyze over-dispersed, under-dispersed, and equi-dispersed data sets.

- As the value of $\theta$ rises, the CV tends to increase.

Table 1. Descriptive measures for the different combination of the parameters of DT distribution.

| Parameter | Descriptive Measures | | | | | |
|---|---|---|---|---|---|---|
| $\theta$ | Mean | Variance | Skewness | Kurtosis | IOD | CV |
| 1.001 | 98.8320 | 8.2885 | -20.5690 | 469.1044 | 0.0838 | 0.0291 |
| 1.005 | 94.54885 | 199.106 | -3.6319 | 13.3240 | 2.1058 | 0.1492 |
| 1.010 | 81.4812 | 575.2569 | -1.2399 | 0.4283 | 7.0599 | 0.2943 |
| 1.050 | 19.9959 | 81.0311 | 0.2090 | -0.4210 | 4.0523 | 0.4501 |
| 1.100 | 9.9920 | 21.2957 | 0.2081 | -0.4187 | 2.1312 | 0.4618 |
| 1.248 | 4.0301 | 4.0376 | 0.2035 | -0.4075 | 1.0018 | 0.4985 |
| 1.750 | 1.2866 | 0.6969 | 0.1956 | -0.4256 | 0.5416 | 0.6488 |
| 2.000 | 0.9422 | 0.4820 | 0.2086 | -0.5048 | 0.5115 | 0.7368 |
| 2.500 | 0.5906 | 0.3074 | 0.2159 | -0.9031 | 0.5205 | 0.9387 |

### 3.4 Entropy

Entropy is a crucial measure of uncertainty and has many applications in applied fields. One of the important entropy is Rényi entropy (RE) (see, Rényi, 1961). For the DT distribution, the RE can be defined as ($\rho > 0, \rho \neq 1$)

$$I_R(\rho) = \frac{1}{1-\rho} \log \sum_{y=0}^{\infty} p_y^{\rho}$$

$$= \frac{1}{1-\rho}\left(\rho + \log \sum_{x=0}^{\infty} \theta^{\rho y}(\exp(-\theta^y) - \theta \exp(-\theta^{(y+1)}))^{\rho}\right).$$

### 3.5 Survival and hazard rate functions

The SF and hazard rate function (HRF) of the DT distribution is respectively given by,

$$S(y;\theta) = P(Y > y) = \theta^{y+1} \exp(1-\theta^{y+1}); y = 0,1,2,...,$$

$$H(y;\theta) = P(Y = y | Y \geq y) = 1 - \theta \exp(\theta^y - \theta^{(y+1)}); y = 0,1,2,....$$

Figure 2 depicts various plots of HRF of the proposed model. From the HRF plot, it is easily visible that the HRF of the DT distribution is increasing. Also, $\lim_{y \to \infty} H(y;\theta) = \lim_{\theta \to \infty} H(y;\theta) = \lim_{\theta \to 1} H(y;\theta) = 1$. The reversed hazard rate function (RHRF) and the second rate of failure (SRF) of the proposed model are

$$H^*(y;\theta) = P(Y = y \mid Y \le y) = \frac{\exp(1)\theta^y(\exp(-\theta^y) - \theta\exp(-\theta^{(y+1)}))}{1 - \theta^{y+1}\exp(1 - \theta^{(y+1)})}; y = 0,1,2,\ldots, \text{ and}$$

$$H^{**}(y;\theta) = \log\left[\frac{S(y)}{S(y+1)}\right] = \theta^{y+1}(\theta - 1) - \log\theta; y = 0,1,2,\ldots, \text{ respectively.}$$

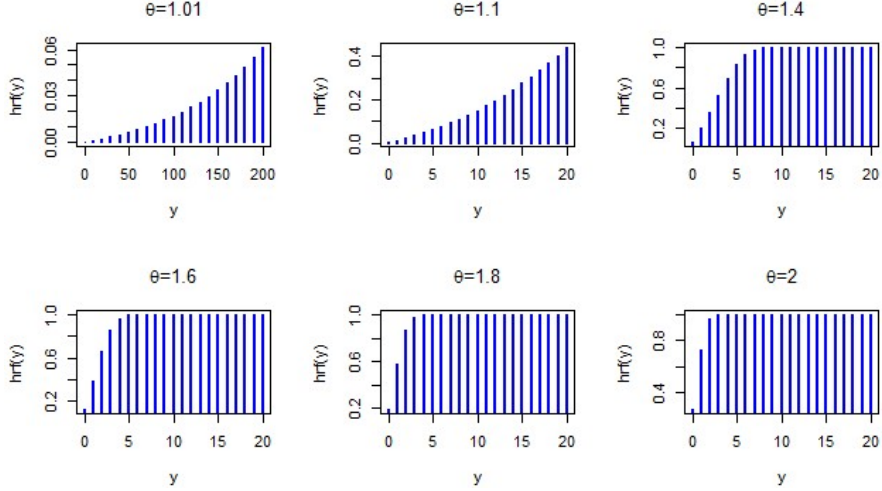

Figure 2. The shapes of HRF of DT distribution for various values of the parameter.

### 3.6 Mean residual lifetime and mean past lifetime

The mean residual life (MRL) function is used extensively in a wide variety of areas, including reliability engineering, survival analysis, and biomedical research since it represents the ageing mechanism. It is well known that the MRL function characterizes the distribution function F uniquely since it contains all of the model's information. In discrete setup, the MRL, symbolized by $m(i)$, can be defined as

$$m(i) = E(Y - i \mid Y \ge i) = \frac{1}{S(i)} \sum_{j=i+1}^{\infty} S(j); i = 0,1,2,\ldots$$

If Y has DT distribution with parameter $\theta$, then the MRL function of Y is

$$m(i) = \frac{1}{\theta^i \exp(-\theta^{i+1})} \sum_{j=i+1}^{\infty} \theta^j \exp(-\theta^{j+1}).$$

The expected inactivity time function or mean past life function (MPL), denoted by $m^*(i)$, measures the time elapsed since the failure of X given that the system has failed sometime before

'i'. It has many applications in a wide variety of areas, including reliability theory and survival analysis, actuarial research, and forensic science. In discrete setup, MPL function is defined as

$$m^*(i) = E(i - X \mid X < i) = \frac{1}{F(i-1)} \sum_{k=1}^{i} F(k-1); \; i = 1, 2, \ldots.$$

By replacing the CDF (6) in the expression of $m^*(i)$, we can easily obtain the MPL for the proposed model.

### 3.7 Stress-strength analysis

The stress-strength (S-S$^*$) analysis is widely applicable in various areas including engineering, medical science, psychology etc. The probability of failure is based on the probability of S exceeding S$^*$. Suppose that the domain of S and S$^*$ is positive, then the stress-strength reliability (R) can be computed as

$$R = P[Y_S \leq Y_{S^*}] = \sum_{y=0}^{\infty} P_{Y_S}(y) S_{Y_{S^*}}.$$

If $Y_S \simeq DT(\theta_1)$ and $Y_{S^*} \simeq DT(\theta_2)$, then $R$ can be expressed as

$$R = \theta_2 \exp(2) \sum_{y=0}^{\infty} (\theta_1 \theta_2)^y \exp(-\theta_2^{y+1}) \left( \exp(-\theta_1^y) - \theta_1 \exp(-\theta_1^{y+1}) \right). \quad (12)$$

Given the difficulty of obtaining an explicit expression for $R$ in this instance, we show this feature quantitatively using the R software. Tables 2 illustrates the calculated values of $R$ for various parameter combinations. From Table 2, we infer that for a fixed value of $\theta_2$, reliability increases as $\theta_1$ increases, whereas for the particular value of $\theta_1$, $R \to 0$ as $\theta \to \infty$.

Table 2. The numerical values of $R$ for fixed values of $\theta_1$ and $\theta_2$.

| Parameter | | $\theta_2$ | | | | |
|---|---|---|---|---|---|---|
| | | 1.001 | 1.010 | 1.050 | 1.250 | 1.500 |
| $\theta_1$ | 1.001 | 0.02093 | 0.00617 | 0.00024 | 0.00001 | 0.00000 |
| | 1.010 | 0.98078 | 0.49681 | 0.02554 | 0.00100 | 0.00026 |
| | 1.050 | 0.99974 | 0.97225 | 0.48476 | 0.02531 | 0.00636 |
| | 1.250 | 0.99999 | 0.99855 | 0.96300 | 0.43096 | 0.13811 |
| | 1.500 | 0.99999 | 0.99950 | 0.98739 | 0.73634 | 0.37738 |

## 3.7 Order statistics

Let $Y_1, Y_2, ..., Y_n$ be a random sample (RS) from the $DT(\theta)$ distribution. Also, let $Y_{(1)}, Y_{(2)}, ..., Y_{(n)}$ represents the corresponding order statistics. Then the CDF of the $r^{th}$ order statistic say $W = Y_{(r)}$ is given by

$$F_r(w) = \sum_{i=r}^{n} \binom{n}{i} F^i(w).[1-F(w)]^{n-i},$$

$$= \sum_{i=r}^{n} \sum_{k=0}^{n-i} (-1)^k \binom{n}{i}\binom{n-i}{k} (1-\theta^{(w+1)} \exp(1-\theta^{(w+1)}))^{i+k}. \tag{13}$$

The corresponding PMF of $r^{th}$ order statistics is

$$f_r(w) = F_r(w) - F_r(w-1),$$

$$= \sum_{i=r}^{n} \sum_{k=0}^{n-i} (-1)^k \binom{n}{i}\binom{n-i}{k} \{(1-\theta^{w+1} \exp(1-\theta^{(w+1)}))^{i+k} - (1-\theta^w \exp(1-\theta^w))^{i+k}\}. \tag{14}$$

Particularly, by setting $r=1$ and $r=n$ in Equation (14), we can obtain the PMF of minimum $(\{Y_{(1)}, ..., Y_{(n)}\})$ and the PMF of maximum $(\{Y_{(1)}, ..., Y_{(n)}\})$, respectively.

## 3.8 Infinite divisibility

In this section, the property of infinite divisibility of the DT distribution is examined. This property is critical in the theorems of probability theory, modelling problems, and waiting time distribution. A probability distribution with PMF $p_x$, $x = 0,1,2,...$ is infinite divisible if $p_x \leq e^{-1} \ \forall \ x = 1,2,...$ [Steutel and Van Harn, 2006]. For DT distribution with $\theta = 2$, we observe that $p_1 = 0.5366$ which is greater than $e^{-1} (= 0.3679)$. Hence in general, DT distribution is not infinitely divisible. Further, since the classes of self-decomposition and stable distributions, in their discrete concepts, are subclasses of infinitely divisible distributions, therefore a DT distribution can neither be self-decomposable nor stable in general.

## 4. Classical Estimation

In this section, we address the problem of estimation through the method of maximum likelihood and the method of moment estimation.

### 4.1. Method of maximum likelihood (MLE)

Let $Y_1, Y_2, ..., Y_n$ be a random sample (RS) of size $n$ with mean $\bar{y}$, then the likelihood-function (LF) for DT distribution can be written as

$$L(\underset{\sim}{y},\theta) = \exp(n)\theta^{n\bar{y}} \prod_{i=1}^{n} (\exp(-\theta^{y_i}) - \theta \exp(-\theta^{(y_i+1)})). \qquad (15)$$

The log-likelihood (LL) function can be represented as

$$\log L(\underset{\sim}{y},\theta) = n + n\bar{y}\log\theta + \sum_{i=1}^{n} \log(\exp(-\theta^{y_i}) - \theta\exp(-\theta^{(y_i+1)})). \qquad (16)$$

Taking the partial derivative of the LL function with respect to the parameter, we get the following normal-equation,

$$\frac{\partial \log L}{\partial \theta} = \frac{n\bar{y}}{\theta} + \sum_{i=1}^{n} \frac{E_1 E_2 - y_i \theta^{y_i - 1}}{1 - \theta E_1} = 0, \qquad (17)$$

where $E_1 = \exp(\theta^{y_i} - \theta^{y_i+1})$ and $E_2 = (y_i + 1)\theta^{y_i+1} - 1$.

The ML estimator of $\theta$ can be found by simplifying Equation (17), but unfortunately, this equation does not yield an analytical solution. Therefore, we use an iterative approach such as Newton-Raphson (NR) to calculate the estimate computationally.

**4.2 Method of moment estimation**

In this estimation process, firstly, we equate population moment(s) to the corresponding sample moment(s) and then solve this equation for the unknown parameter(s). In our case, the concerned equation is

$$\bar{y} = \sum_{i=1}^{\infty} \theta^i \exp(1 - \theta^i), \qquad (18)$$

where $\bar{y}$ represents the mean based on the RS $y_1, y_2, ..., y_n$ drawn from the DT distribution (5). We can obtain the MOM estimator $\hat{\theta}_{MOM}$, by solving Equation (18) for $\theta$. Since Equation (18) does not provide the MOM estimator of $\theta$ in explicit form, so we can use numerical methods to compute $\hat{\theta}_{MOM}$.

# 5. The real data application

In this part, we use two real data sets to demonstrate the relevance of the DT distribution. The fitting capability of the proposed model has been compared to the models listed in Table 3.

Table 3. The competitive models.

| Model | Parameter(s) | Abbreviation | Reference(s) |
|---|---|---|---|
| Geometric | $\theta$ | Geo | - |
| Discrete Lindley | $\alpha$ | DsLi | G´omez-D´eniz and Calder´ın-Ojeda (2011) |
| Discrete Rayleigh | $\theta$ | DR | Roy (2004) |
| Discrete Poisson Lindley | $\alpha$ | DPL | Sankaran (1970) |
| Discrete Burr | $(\alpha, \beta)$ | DBr | Krishna and Pundir (2009) |
| Discrete Pareto | $\theta$ | DPa | Krishna and Pundir (2009) |
| Two Parameter Discrete Half Logistic | $(\alpha, \beta)$ | DHLo-II | El-Morshedy et al. (2021c) |
| Discrete Perks | $(\alpha, \beta)$ | DP | Tyagi et al. (2020) |
| Discrete Weibull | $(q, \beta)$ | DW | Nakagawa and Osaki (1975) |
| Discrete Logistic | $(\alpha, \beta)$ | DLOG | Chakraborty and Chakravarty (2016) |
| A Flexible discrete model with one parameter | $\alpha$ | DsFx-I | Eliwa and El-Morshedy (2021) |
| Poisson Bilal distribution | $\theta$ | PB | Altun (2020) |

For comparison purposes, the estimation of the fitted models has been done through ML estimation. The model comparison has been done through the following measures:

- ❖ –LL,
- ❖ Akaike information criterion (AIC),
- ❖ corrected Akaike information criterion (CAIC),
- ❖ Bayesian information criterion (BIC),
- ❖ Kolmogorov-Smirnov (KS) statistics with the associated p-value.

Here, the lower value of these criteria except the p-value and the higher p-value indicates the best fit. All required computations have been done using open-source software R.

*The first data set (I)*: In the first application, we consider the daily new cases in India from 16 March 2021 to 08 April 2021. The data is available at https://www.worldometers.info/coronavirus/country/india-sar/. The original data values are

28869, 35838, 39643, 40950, 43815, 40611, 47264, 53419, 59069, 62291, 62631, 68206, 56119, 53158, 72182, 81441, 89019, 92998, 103793, 96557, 115269, 126315, 131893, 14482.

This data set is modelled with DT and other competitive models listed in Table 3. For ease of fitting, data have been divided by 10,000 and their floor values have been stored. Table 4 contains the estimated parameters and their corresponding standard errors (SEs) as well as the various fitting measures discussed earlier. From Table 4, we conclude that the DT model is the best-performed model among others since it has the lowest values of AIC, BIC, CAIC, HQIC, and K-S test statistics with the highest p-value. We have plotted the –LL and CDF plots in Figure 3 (upper left and upper right panel). This figure not only confirms the unique existence of the MLE but also portrays that the fitted CDF closely follow the pattern of the empirical CDF for the considered data.

Table 4. The ML estimate (SE) and various goodness of fit measures under data set I.

| Model | MLE(SE) | -LL | AIC | BIC | CAIC | K-S | P-value |
|---|---|---|---|---|---|---|---|
| DT | 1.1447 (0.0121) | 61.6498 | 125.2997 | 126.4778 | 125.4815 | 0.12640 | 0.8374 |
| DW | 0.0035(0.0034), 2.5990 (0.4083) | 61.1957 | 126.3914 | 128.7475 | 126.9628 | 0.12669 | 0.7907 |
| DR | 0.9844(0.0031) | 61.8800 | 125.76 | 126.9381 | 125.9418 | 0.15186 | 0.6373 |
| DP | 0.0252(0.0208), 0.5020(0.0974) | 62.2001 | 128.4003 | 130.7564 | 128.9717 | 0.13958 | 0.6869 |
| DLOG | 0.5860(0.05317), 7.4515(0.6792) | 62.7109 | 129.4219 | 131.778 | 129.9934 | 0.13598 | 0.7166 |
| PB | 0.1136(0.0187) | 67.2632 | 136.5266 | 137.7046 | 136.7084 | 0.30641 | 0.0169 |
| DsLi | 0.7920(0.0269) | 67.8623 | 137.7246 | 138.9027 | 137.9064 | 0.31724 | 0.0120 |
| DPL | 0.2460(0.0396) | 68.7832 | 139.5665 | 140.7445 | 139.7483 | 0.32841 | 0.0083 |
| DHLo-II | 0.8548(0.0316), 0.7729(0.0626) | 69.1321 | 142.2644 | 144.6205 | 142.8358 | 0.35068 | 0.0038 |
| DsFx-I | 0.9020(0.0167) | 71.0351 | 144.0703 | 145.2484 | 144.2521 | 0.81633 | <0.0001 |
| Geo | 0.8716(0.0244) | 71.6627 | 145.3255 | 146.5035 | 145.5073 | 0.29772 | 0.0284 |
| DB | 0.9261(0.0709), 6.6364(6.4802) | 87.3628 | 178.7273 | 181.0834 | 179.2987 | 0.74374 | <0.0001 |
| DPa | 0.6217(0.0603) | 92.3961 | 186.7923 | 187.9703 | 186.9741 | 0.72802 | <0.0001 |

Table 5 consists of MLE and MOM estimates with their SEs. To compare these methods, the KS statistics with associated p-values are also provided in Table 5. From Table 5, we can easily observe the MLE perform better as compared to MOM since MLE has lesser K-S statistics, SE and higher p-value.

Table 5. The different estimate, SE, and with K-S and p-value under data set I.

| Method | Estimate | SE | K-S | P-value |
|---|---|---|---|---|
| MLE | 1.1447 | 0.0121 | 0.1264 | 0.8374 |
| MOM | 1.1372 | 0.0367 | 0.1544 | 0.6162 |

*The second data set (II)*: This data set gives the survival times of a group of laboratory mice, which were exposed to a fixed dose of radiation at an age of 5 to 6 weeks [Lawless, 2011, pp. 445]. This group of mice lived in a conventional lab environment. The cause of death for each mouse was assigned after autopsy to be one of three things: thymic lymphoma (C1), reticulum cell sarcoma (C2), or other causes (C3). Here, we have used the data set under C3 only. The mice are all died by the end of the experiment, so there is no censoring. The data values are:

40, 42, 51, 62, 163, 179, 206, 222, 228, 252, 259, 282, 324, 333, 341, 366, 385, 407, 420, 431, 441, 461, 462, 482, 517, 517, 524, 564, 567, 586, 619, 620, 621, 622, 647, 651, 686, 761, 763.

The above data set is modelled with DT and DW, DR, PB, DsLi, DPL, Geo, DB, DPa models. The estimated parameters and other fitting measures are reported in Table 6. From the outcomes of Table 6, we conclude that the DT distribution is the best choice among other competitive models since it has the lowest values of –LL, AIC, BIC, CAIC, HQIC, and K-S statistics with the highest P-value. Figure 3 (lower left and lower right panel) also depicts that DT distribution has a unique MLE for the given data and it is well enough to model this data.

Table 6. The MLE (SE) and various goodness of fit measures under data set II.

| Model | MLE (SE) | -LL | AIC | BIC | CAIC | K-S | P-Value |
|---|---|---|---|---|---|---|---|
| DT | 1.0024(0.0002) | 262.0291 | 526.0581 | 527.7217 | 526.1663 | 0.0907 | 0.9049 |
| DW | 0.9999 (3.548e-07), 2.0772 (0.0318) | 263.1519 | 530.3039 | 533.6310 | 530.6372 | 0.1008 | 0.8223 |
| DR | 0.9999 (5.874e-07) | 263.1909 | 528.3818 | 530.0454 | 528.4899 | 0.1080 | 0.7525 |
| PB | 0.0020(0.0002) | 267.1738 | 536.3476 | 538.0112 | 536.4557 | 0.1597 | 0.2723 |
| DsLi | 0.9951(0.0005) | 266.9048 | 535.8097 | 537.4733 | 535.9178 | 0.1587 | 0.2797 |
| DPL | 0.0048(0.0005) | 266.9121 | 535.8242 | 537.4878 | 535.9323 | 0.1588 | 0.2786 |
| Geo | 0.9975(0.0004) | 273.9544 | 549.9088 | 551.5723 | 550.0169 | 0.2385 | 0.0236 |
| DB | 0.9282(0.0621), 2.3077(2.0575) | 334.6387 | 673.2775 | 676.6046 | 673.6108 | 0.6803 | <0.0001 |
| DPa | 0.8422(0.0231) | 334.8421 | 671.6843 | 673.3478 | 671.7924 | 0.6801 | <0.0001 |

Table 7 consists of MLE and MOM estimates with their SEs, and K-S statistics with associated p-values. From Table 7, we can easily observe the MLE perform better as compared to MOM since MLE have lesser K-S statistics, SE and higher p-value.

Table 7. The different estimates, SE and with K-S and p-value under data set II.

| Method | Estimate | SE | K-S | P-Value |
|--------|----------|--------|--------|---------|
| MLE    | 1.0024   | 0.0002 | 0.0907 | 0.9049  |
| MOM    | 1.0096   | 0.0022 | 0.2047 | 0.0759  |

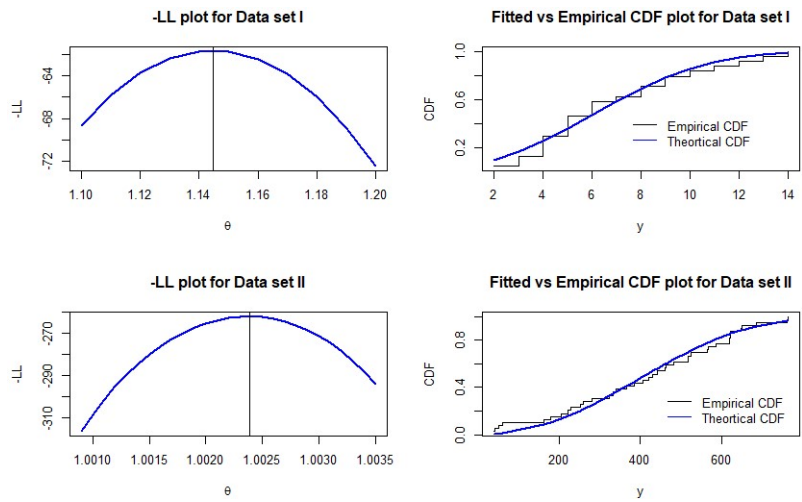

Figure 3. The –LL and CDFs plots for data set I and II.

## 6. Conclusion

In this article, a new one-parameter discrete distribution so-called discrete Teissier distribution is obtained. Its several impressive features have been discussed. The classical estimation using the method of maximum likelihood and method of moment is performed. Finally, the fitting capability of the proposed model using two real data sets is demonstrated. In future, we will develop its bivariate extension.

**Acknowledgement**

The last author is thankful to the Department of Science and Technology, India for providing financial aid for the research work under the Inspire Fellowship Program vide letter number DST/INSPIRE Fellowship/2017/IF170038.